\documentclass[12pt,a4paper]{article}
\usepackage{graphicx}
\usepackage{wrapfig,amsmath}
\usepackage{amssymb}

\newcommand{\ri}{{\mathrm{i}}}
\newcommand{\order}[1]{\ensuremath{{\cal O}(#1)}}

\newcommand{\GeV}{\ensuremath{\unskip\,\mathrm{GeV}}}

\newcommand{\fb}{\unskip\,\mathrm{fb}}

\newcommand{\msbar}{\ensuremath{\overline{\text{MS}}}}
\newcommand{\alphas}{\ensuremath{\alpha_s}}
\newcommand{\mt}{\ensuremath{m_t}}
\newcommand{\mgl}{\ensuremath{m_{\tilde{g}}}}
\newcommand{\mhpm}{\ensuremath{m_{H^{\pm}}}}

\newcommand{\dmb}{\ensuremath{\Delta m_b}}
\newcommand{\mbb}{\ensuremath{\overline{m}_b}}
\newcommand{\mbmb}{\ensuremath{\overline{m}_b(\overline{m}_b)}}

\newcommand{\tabre}[1]{Table~\ref{#1}}
\newcommand{\figre}[1]{Fig.~\ref{#1}}

\newcommand{\fa}{\texttt{FeynArts}}
\newcommand{\fc}{\texttt{FormCalc}}
\newcommand{\lt}{\texttt{LoopTools}}
\newcommand{\vegas}{\texttt{Vegas}}

\begin{document}

\begin{titlepage}

\rightline{DESY 10-153}
\rightline{HD-THEP-10-17}
\rightline{MPP-2010-126}

\begin{center}
\boldmath
{\large{\bf Supersymmetric QCD corrections to $e^+e^-\to t\bar{b}H^-$ and the
Bernstein-Tkachov method of loop integration}}
\unboldmath

\vspace*{4mm}

B.~A.~Kniehl$^1$, M.~Maniatis$^2$ and M.~M.~Weber$^3$

\vspace{4mm}
$^1${\it II. Institut f\"ur Theoretische Physik, Universit\"at Hamburg,\\
  Luruper Chaussee 149, 22761 Hamburg, Germany}\\[2mm]

$^2${\it Institut f\"ur Theoretische Physik, Universit\"at Heidelberg,\\
Philosophenweg 16, 69120 Heidelberg, Germany}\\[2mm]

$^3${\it Max-Planck-Institut f\"ur Physik (Werner-Heisenberg-Institut),\\
F\"ohringer Ring 6, 80805 M\"unchen, Germany}
\end{center}

\begin{abstract}
The discovery of charged Higgs bosons is of particular importance,
since their existence is predicted by supersymmetry and they are absent in
the Standard Model (SM).  If the charged Higgs bosons are too heavy to
be produced in pairs at future linear colliders, single production
associated with a top and a bottom quark is enhanced in parts of the
parameter space.  We present the next-to-leading-order
calculation in supersymmetric QCD within the minimal supersymmetric SM (MSSM),
completing a previous calculation of the SM-QCD corrections.  In
addition to the usual approach to perform the loop integration
analytically, we apply a numerical approach based on the
Bernstein-Tkachov theorem. In this framework, we avoid some of the
generic problems connected with the analytical method.
\end{abstract}

\end{titlepage}


%
%
\section{Introduction}
\label{sectionintro}

The discovery of charged Higgs bosons~($H^{\pm}$) would be instant
evidence for physics beyond the SM, which only accommodates
a single neutral Higgs boson.  Charged Higgs bosons appear
in models with two Higgs doublets, as are required by
supersymmetric (SUSY) extensions of the SM. For this reason, there
is much interest in charged-Higgs-boson physics (review articles
on this subject include, for instance, Refs.~\cite{Roy:2004az, Djouadi:1992pu,
Kanemura:2000cw}).

Phenomenologically, the Large Hadron Collider~(LHC) will be the first
collider with the potential to discover the $H^{\pm}$ bosons.  If the
charged-Higgs-boson mass $m_H^{\pm}$ is not too large, i.e.\ if
$m_H^{\pm} < m_t-m_b$,
the dominant production channel is via top-quark pair production
$gg \rightarrow t \bar{t}$ with subsequent
decay of a top quark or antiquark into a charged Higgs boson,
$t \rightarrow b H^+$ or $\bar{t} \rightarrow \bar{b} H^-$, respectively.
If the $H^{\pm}$ bosons are too heavy for this subsequent top-quark decay,
then the dominant production channel would be
bottom-gluon fusion, $gb \rightarrow t H^-$ and $g\bar{b}
\rightarrow \bar{t} H^+$
\cite{Zhu:2001nt,Gao:2002is,Plehn:2002vy,Berger:2003sm,Kidonakis:2005hc}.
In all
these processes, the charged-Higgs-boson signal has to be carefully
separated from large SM-QCD background at hadron colliders.

Despite the fact that charged Higgs bosons may be discovered at the
LHC, a precise determination of their properties will only be possible at a
linear collider, such as the proposed international linear collider
(ILC).  If the value of $\mhpm$ is not too large, i.e.\ if
$m_{H^{\pm}} < \sqrt{s}/2$, the dominant production channel will be
$e^+e^- \rightarrow H^+H^-$
\cite{Komamiya:1988rs,Guasch:2001hk}. On the other hand, this production
channel
may not be accessible at the ILC because of the limited center-of-mass
energy $\sqrt{s}$. In this case, charged Higgs bosons may be copiously produced
singly via the two channels \cite{Kanemura:2000cw}
\begin{eqnarray}
e^+  e^- & \rightarrow & \tau^+  \nu_{\tau}  H^- + \text{c.c.}
\label{singleHiggstau}
\\
e^+  e^- & \rightarrow & t  \bar{b}  H^- + \text{c.c. .}
\label{singleHiggstop}
\end{eqnarray}
The first production channel (\ref{singleHiggstau}) was investigated in
Ref.~\cite{Kanemura:2000cw}, where single
charged-Higgs-boson production processes were systematically compared with
each other
with leading-order (LO) accuracy.  Here, we focus on the second production
channel (\ref{singleHiggstop}), which is enhanced in parts of the
parameter space.  Since QCD corrections are typically large, we present
a computation with next-to-leading-order (NLO) accuracy,
to ${\cal O}(\alpha_s)$.

One part of the NLO calculation consists of the
SM-QCD contribution, i.e.\ the purely gluonic corrections, which were
already presented in Ref.~\cite{Kniehl:2002zz}.  Here, we add the
SUSY-QCD contribution, due to squark and gluino loops,
in the minimal SUSY SM~(MSSM)~\cite{Fayet:1976cr, Nilles:1983ge, Haber:1984rc}.
Since the MSSM makes no assumption about the SUSY-breaking
mechanism, but just uses explicit SUSY-breaking terms in the
Lagrangian, it may be considered as representative for a wide
class of SUSY models. The SM-QCD and SUSY-QCD
parts taken together yield a complete prediction with ${\cal O}(\alpha_s)$
accuracy.

In the calculation of the virtual corrections, we encounter loop integrals.
Using the conventional analytic approach of Ref.~\cite{'tHooft:1978xw}, all
scalar loop integrals can be expressed in terms of logarithms and
dilogarithms.  Furthermore, using the reduction algorithm of
Ref.~\cite{Passarino:1978jh}, all tensor loop integrals, i.e.\ integrals
containing loop momenta in the numerator, can be expressed in terms of
scalar integrals. Therefore, a full analytic solution for one-loop
integrals exists. However, in general, this approach has a number of drawbacks.
First of all, the number of
dilogarithms in the analytic expression of a scalar integral increases
rapidly with an increasing number of external legs. This may lead to cancellations for multileg integrals in
certain kinematic regions \cite{vanOldenborgh:1989wn}.
Furthermore the tensor reduction of Ref.~\cite{Passarino:1978jh} introduces
inverse Gram determinants. These may vanish at the phase-space
boundary, even though the tensor coefficients themselves remain regular there.
There may thus be cancellations among terms in the
numerator, which may lead to numerical instabilities.
While these generic problems may still be overcome in the case under
consideration here, where at most box diagrams occur, by exercising care in
the phase-space integrations, they become quite severe for five and more
external legs.
To address these problems, several improved reduction algorithms have been
constructed \cite{Denner:2005nn, Binoth:2005ff, Fleischer:2010mq,
  Diakonidis:2008ij} allowing a numerically stable evaluation of the tensor
integral coefficients.  A different solution, typically used in the context of
nondiagrammatic methods to calculate loop amplitudes, is to employ high-precision
arithmetics for potentially unstable phase-space points (see
Refs.~\cite{Berger:2009zb, Bern:2007dw, Bern:2008ef} for reviews of these
techniques).  However, these improvements come at the price of either increased
complexity of the more elaborate reduction algorithms or increased runtime from
the high-precision evaluations.

Finally, within dimensional regularization in $D=4-2\epsilon$ space-time
dimensions, the evaluation of the
loop-by-loop contribution to a two-loop correction makes it necessary
to expand the one-loop integrals beyond the constant term in the
expansion about $D=4$ dimensions. An analytic calculation of these higher-order
terms is rather complicated.

We explore here an alternative numerical approach to the evaluation of one-loop
tensor and scalar integrals.  This strategy is described in detail in
Ref.~\cite{Ferroglia:2002mz} and is based on the Bernstein-Tkachov (BT) theorem
\cite{Tkachov:1996wh}, which can be used to rewrite one-loop integrals
in Feynman-parametric representation in a form better suited for
numerical evaluation.  This technique yields a fast and reliable
numerical calculation of multileg one-loop integrals.  This method is
numerically stable also for exceptional momentum configurations and
easily allows for the introduction of complex masses and the calculation
of higher orders of the expansion about $D=4$ dimensions.
Numerical methods based on the BT theorem have also been developed for two-loop
self-energy and vertex integrals \cite{Passarino:2001jd, Ferroglia:2003yj,
  Actis:2004bp, Passarino:2006gv}, and found several applications
\cite{Hollik:2005va, Passarino:2007fp, Actis:2008ug, Actis:2008ts}.
While these applications involved two-loop integrals with full mass dependence,
the phase-space integrations were trivial.
We adopt the BT method in our calculation and compare it to the conventional
analytic approach in order to investigate its performance in the computation of
cross sections including nontrivial phase-space integrations at one loop.



%
%

\section{The calculation}
\label{sectioncalculation}

\begin{figure}
\centerline{\includegraphics[bb= 160 583 454 652, width=0.7\textwidth]{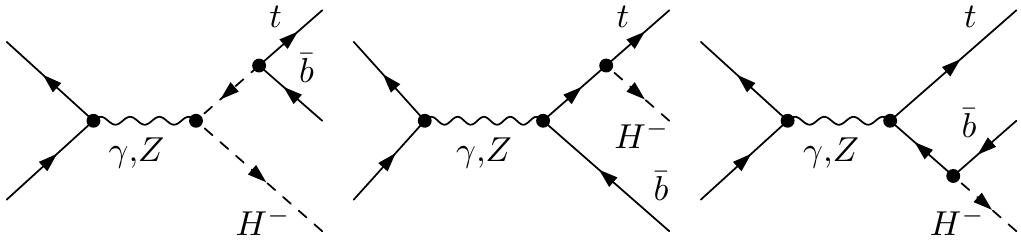}}
\caption{\em Feynman diagrams of the process $e^+ e^- \rightarrow t \bar{b} H^-$ at
  Born level. Those of the charge-conjugated process
$e^+ e^- \rightarrow b \bar{t} H^+$ are not shown.}
\label{figBorn}
\end{figure}

\begin{figure}
\centerline{\includegraphics[bb= 99 585 516 668, width=1\textwidth]{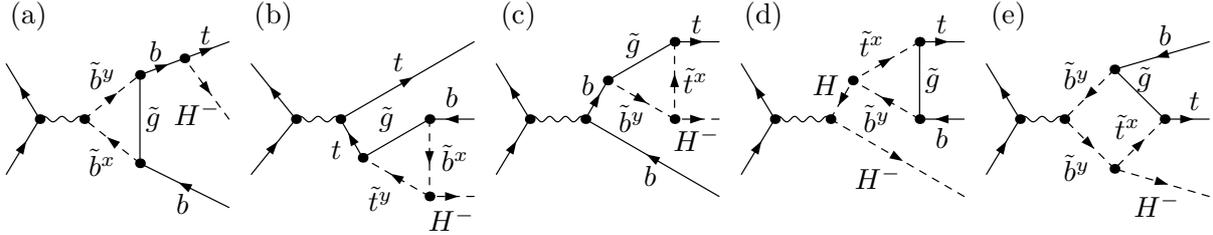}}
\caption{\em Representative one-loop diagrams of the SUSY-QCD corrections to
  the process $e^+ e^- \rightarrow t \bar{b} H^-$.  The inner lines represent
  gluinos~($\tilde{g}$) as well as the various top ($\tilde t^x$) and bottom
($\tilde b^x$) squarks with $x=1,2$ labeling the mass eigenstates.}
\label{diags-virt}
\end{figure}

We consider the process $e^+ e^- \rightarrow t \bar{b} H^-$ and its
charge-conjugate counter part $e^+ e^- \rightarrow b \bar{t} H^+$. The
corresponding LO
Feynman diagrams are shown in Fig.~\ref{figBorn}.  Since the cross sections for
both processes are identical due to CP invariance, we only show results for the
$t\bar{b}H^-$ final state in the following.  Furthermore, we only
consider the kinematical regime where neither the intermediate Higgs bosons
nor the top
quark can be on shell and, therefore, we do not have to include
the respective finite widths.

The SUSY-QCD corrections arise from loop contributions containing gluinos,
stops, and sbottoms in the propagators; see \figre{diags-virt} for some
 representative
diagrams. This virtual contribution may be classified into two-, three-, and
four-point
integrals. Since the particles in the loops are massive, no infrared or collinear
singularities arise and no real-emission contribution has to be included.

The calculation is performed with the help of the program packages
\fa\ \cite{Kublbeck:1990xc} and \fc\ \cite{Hahn:1998yk}.  The analytical
expressions from \fc\ are post-processed and translated to a C++ code.
The one-loop
integrals are evaluated using the BT method as described in
detail below.  We perform a second calculation using analytical loop
integrals as implemented by \lt\ \cite{Hahn:1998yk}.  This calculation is also
based on \fa\ and \fc. The phase-space integration is performed with the
Monte Carlo integration routine \vegas\ \cite{Lepage:1977sw} in both cases.

The renormalization of the strong-coupling constant $\alphas$ is performed in
the modified minimal-subtraction ($\msbar$) scheme of dimensional
regularization with the squarks and gluinos decoupled from the running.
The top
quark only contributes to the running of $\alphas$ above the scale $\mt$, by
increasing the number of active quark flavors from $n_f=5$ to $n_f=6$. The
quark masses and wave functions are renormalized on shell, including the top-quark
mass in the Yukawa coupling.  For the bottom-quark mass in the Yukawa coupling,
 we use the
running QCD $\msbar$ version $\mbb(\mu)$, with~$\mu$ being
the renormalization scale, and optionally perform a resummation of
large SUSY corrections as described below.  We use the running bottom-quark
mass, since the
pure QCD corrections contain large logarithms of the type $\log(\mu/m_b)$ originating from the
Yukawa interaction. These can be resummed by using the running bottom-quark mass
$\mbb(\mu)$ \cite{Braaten:1980yq, Drees:1990dq}. The calculation of the
purely gluonic QCD corrections showed that the
bulk of the QCD corrections may be absorbed by using the running bottom-quark
mass \cite{Kniehl:2002zz}.
Since the squark and gluino masses enter only at NLO we do not need to
renormalize them.

In the MSSM, two Higgs-boson doublets, denoted by
$\hat{H}_u=(H_u^+,H_u^0)$ and $\hat{H}_d=(H_d^0,H_d^-)$, are required.
The $H_u^0$ field
is responsible for the generation of the up-type-fermion masses and the
$H_d^0$ field for the down-type-fermion masses, i.e.\ the bottom quark couples
to $H_d^0$ but not to $H_u^0$. Nevertheless, the coupling of the
bottom quark to the $H_u^0$ field is dynamically generated via loops
\cite{Carena:1999py}. Although this coupling is loop suppressed, once
the $H_{u/d}^0$ fields acquire their vacuum expectation values $v_{u/d}$, a
large value of $v_u$ may compensate a small loop contribution, i.e.\
these effects may be considerable for large values of
$\tan \beta = v_u/v_d$.
These large $\tan \beta$-enhanced contributions may be resummed to all orders
\cite{Carena:1999py} by replacing the bottom-quark mass in the Yukawa coupling
as
\begin{equation}
m_b \to \frac{\overline{m}_b(\mu)}{1 + \dmb} \left( 1 - \frac{\dmb}{\tan^2\beta}
  \right) ,
\end{equation}
where
\begin{align}
  \dmb &= \frac{2\alphas(\mu)}{3\pi} \mgl \tilde{\mu} \tan\beta\;
  I(\mgl, m_{\tilde{b}_1},  m_{\tilde{b}_2}),  \nonumber \\
I(a,b,c) &= \frac{1}{(a^2 - b^2) (b^2 - c^2) (a^2 - c^2)}
\left( a^2 b^2 \log\frac{a^2}{b^2} + b^2 c^2 \log\frac{b^2}{c^2}
+ c^2 a^2 \log\frac{c^2}{a^2} \right).
\label{eq:deltamb}
\end{align}
Here, $\tilde{\mu}$ is the Higgs-Higgsino mass parameter of the superpotential.
In order to prevent double counting, an extra counterterm of the form
\begin{equation}
\delta m_b^{\text{Yuk}} = \mbb(\mu) \dmb \left( 1 + \frac{1}{\tan^2\beta} \right)
\end{equation}
for the Yukawa coupling is needed at NLO. The resummation
formalism was extended to also include the dominant terms in the trilinear
coupling $A_b$ \cite{Guasch:2003cv}. Since these contributions are small for
our parameter values, we do not include them in the resummation.

\subsection{Numerical evaluation of loop integrals}

Within dimensional regularization, any
scalar one-loop integral can be expressed as an integral over
Feynman parameters as
\begin{equation}\begin{split}
I_N^D & =\frac{(2\pi\mu)^{4-D}}{\ri\pi^2} \int \text{d}^Dq
  \frac{1}{[q^2-m_1^2] [(q+p_1)^2-m_2^2] \cdots [(q+p_{N-1})^2-m_N^2]}\\[1ex]
  & = (4\pi\mu^2)^\epsilon\; \Gamma(N-2+\epsilon) (-1)^N
    \int \text{d}S_{N-1} V(x_i)^{-(N-2+\epsilon)},
\label{eq:scal-int-def}
\end{split}\end{equation}
where
\[ \int \mathrm{dS}_n \; = \int_0^1 \mathrm{d}x_1\;
    \int_0^{x_1} \mathrm{d}x_2\; \cdots \int_0^{x_{n-1}} \mathrm{d}x_n
\]
and $V$ is a quadratic form in the $N-1$ Feynman parameters $x_i$,
\[V(x) = x^T H x + 2K^Tx + L - \ri\delta.\]
The coefficients $H$, $K$, and $L$ of $V$ are given in terms of the
momenta $p_i$ and the masses $m_i$.

In general, the quadratic form $V$ can vanish within the integration region,
although, strictly speaking, the zeros are shifted into the complex plane by
the infinitesimal imaginary part $\ri\delta$.
Since the limit $\delta \to 0$ has to be taken in the end, the form
given above is in general not suited for direct numerical integration.

Instead, the integral can be rewritten using the BT theorem
\cite{Tkachov:1996wh} before attempting a numerical
evaluation.
Applied to the case of one-loop integrals, this theorem states that, for any
quadratic form $V(x)$ raised to any real power $\beta$, we have
\begin{equation}
\left[ 1 - \frac{(x-X)_i \partial_i}{2(1+\beta)} \right] V^{1+\beta}(x_i) =
    B \cdot V^\beta(x_i),
\label{eq:bt-theorem}
\end{equation}
where $X = -K^T H^{-1}$, $B = L - K^TH^{-1}K$, and $\partial_i =
\partial/\partial x_i$. Inserting this relation into a Feynman-parameter
integral and integrating by parts, one obtains
\begin{equation}
\int \text{d}S_n V^\beta = \frac{1}{2B(1+\beta)} \left[
    (2+n+2\beta) \int \text{d}S_n V ^{1+\beta}
    - \int \text{d}S_{n-1} \sum_{i=0}^{n} \chi_i V_i^{1+\beta} \right],
\label{eq:bt-1loop}
\end{equation}
where $\chi_i = X_i - X_{i+1}$, with $X_0 = 1$ and $X_{n+1}=0$, and
\[
V_i(x_1,\dots,x_{n-1}) =
\begin{cases}
V(1,x_1,\dots,x_{n-1}) & \text{for } i=0,\\
V(x_1,\dots,x_i,x_i,\dots,x_{n-1}) &\text{for } 0 < i < n,\\
V(x_1,\dots,x_{n-1},0) &\text{for } i=n.
\end{cases}
\]
Applied to the one-loop integral of Eq.~\eqref{eq:scal-int-def}, the first
term inside the brackets in Eq.~\eqref{eq:bt-1loop} corresponds to the
$N$-point integral in $D+2$ dimensions, while the last term is a sum
over $(N-1)$-point integrals in $D$ dimensions obtained by removing
one propagator.

Recursive application of Eq.~\eqref{eq:bt-1loop} allows us to express any
scalar one-loop integral as a linear combination of terms of the form
$\int \mathrm{d}S_k\; V(x_i)^{m-\epsilon}$ with any integer $m\geq0$.
A Taylor expansion up to $\order{\epsilon^a}$ then results in
terms of the form $\int \text{d}S_k \; V^m \cdot \log^{1+a} V$. For
$m=0$, the integrand still contains an integrable (logarithmic)
singularity, while it is smooth for $m>0$. Although larger values of
$m$ lead to smoother integrands, the expressions also grow larger
due to the repeated application of the BT identity \eqref{eq:bt-1loop}.
The optimal choice for $m$ depends on the chosen
numerical integration routine and its ability to deal with integrable
singularities.

The parametric representation of tensor integrals contains Feynman parameters in
the numerator. The procedure
outlined above can also be applied in this case, so that no separate
reduction to scalar integrals is needed. Furthermore, no inverse Gram
determinants are introduced using this approach, making the latter numerically
reliable also for exceptional kinematic configurations.

The BT identity \eqref{eq:bt-1loop} still contains a potentially small
factor of $B$ in the denominator. Although the zeros of $B$ correspond
to the leading Landau singularities, the singular behavior in the
vicinity of this singularity is overestimated by the factor
$1/B$. This may result in numerical cancellations in the
numerator leading to instabilities. We, therefore, use an alternative
BT-like relation for the three-point function \cite{Uccirati:2004vy}, namely
\[
V^{-1-\epsilon}(x) = B^{-\epsilon} \left[ 1 +
  \frac{(x-X)_i \partial_i}{2} \right] \sum_{n=1}^{n=\infty}
\frac{(-\epsilon)^{n-1}}{n!}  \frac{1}{Q(x)}
\ln^n\left(1+\frac{Q(x)}{B}\right),
\]
where $Q(x)$ is defined by the decomposition of $V(x) = Q(x) + B$. For
small values of $B$, the $1/B$ behavior is reduced to $\ln B$, which
is in agreement with the singular behavior near the leading Landau
singularity of a triangle diagram.

We use an implementation of the BT method in \texttt{Mathematica} and C++
\cite{Weber:2004sf} that allows for the calculation of all the appearing
one-loop tensor coefficients.  The Feynman-parameter integration is
performed with a deterministic integration routine, with the number of
integrand evaluations limited to $10^5$.  We verify the results
for the virtual corrections using the BT approach against an analytic
evaluation of the integrals with \lt\ \cite{Hahn:1998yk} for
single points in phase-space and find good agreement within numerical
integration errors.

We could now perform the phase-space integration of the virtual
corrections using Monte Carlo techniques. However, using the BT
approach for the evaluation of the loop integrals requires
a separate numerical integration for each integral at every phase-space
point. This turns out to be very inefficient compared to the
analytic evaluation of the loop integrals.

Therefore, we combine the phase-space integration and the integration over
the Feynman parameters into a single integration that is performed using the
adaptive Monte Carlo integration program \vegas\ \cite{Lepage:1977sw}. The
adaptivity of \vegas\ optimizes the phase-space and Feynman-parameter
integrations at the same time, leading to a significant improvement of the
efficiency. All results for the BT approach shown below are obtained using this
combined integration.



%
%
\section{Results}

Before we present numerical results, we fix the input parameters.
We use the following SM parameters:
\begin{equation} \begin{aligned}
\alpha &= 1/137.0359998, \qquad & \alpha_s(m_Z) &= 0.1184,\\
m_W &= 80.398 \GeV, \qquad  & m_Z &= 91.1876 \GeV, \\
m_t &= 173.1 \GeV, \qquad & \mbmb &= 4.2 \GeV.
\end{aligned} \end{equation}
The running $\msbar$ bottom-quark mass $\mbmb$ is used as input and corresponds
to an
on-shell mass of $m_b^{\text{os}} = 4.58 \GeV$.  As mentioned above, we use the
on-shell bottom-quark mass everywhere, except for the $tbH^{\pm}$ Yukawa
coupling, for which
we always use the running $\msbar$ version.
Both $\mbb(\mu)$ and $\alphas(\mu)$ are evaluated at the renormalization scale
$\mu = (\mhpm + \mt +
m_b)/3$.
The relevant scale for the strong coupling appearing in $\dmb$ in
Eq.~\eqref{eq:deltamb} is $\mu_b = (\mgl + m_{\tilde{b}_1} +
m_{\tilde{b}_2})/3$. However, the two-loop result of
Refs.~\cite{Noth:2008tw,Noth:2010jy,Mihaila:2010mp} for $\dmb$ develops a
maximum at about $\mu_b / 3$ to $\mu_b / 4$, which is, therefore, a more
appropriate scale choice in $\dmb$.  For our choice of Higgs-boson and squark
masses,
$\mu$ accidentally falls into this region. For simplicity, we therefore use
$\mu_b = \mu$ in our calculation.  We note that a shift in the renormalization
scale of $\alpha_s(\mu)$ in Eq.~\eqref{eq:deltamb} formally creates a
contribution beyond the order of our calculation.

\begin{table}
\begin{tabular}{c|ccccc|cccccccc}
&
$m_0$ & $m_{1/2}$ & $A_0$ & $\tan \beta$ & sign~$\mu$
& $m_{H^{\pm}}$ & $m_{\tilde{g}}$
& $m_{\tilde{b}1}$ & $m_{\tilde{b}2}$ & $m_{\tilde{t}1}$ & $m_{\tilde{t}2}$ \\
\hline
SPS1a
&
$100$ & $250$ & $-100$ & $10$ & $+$
& 404.5 & 607.7 & 514.4 & 543.8 & 400.7 & 586.8 \\
SPS4
&
$400$ & $300$ & $0$ & $50$ & $+$
& 343.3 & 734.4 & 617.1 & 682.5 & 548.8 & 698.7
\end{tabular}
\caption{\em Definition of the Snowmass points SPS1a and SPS4. The values of the masses and
  the trilinear coupling $A_0$ are given in units of GeV.  The left part of the
  table shows the mSUGRA parameters the SPS points correspond to
  (the scalar and gaugino mass parameters $m_0$ and $m_{1/2}$, the trilinear
  coupling $A_0$, the ratio of the Higgs vacuum expectation values $\tan \beta$,
  and the sign of the SUSY Higgs-boson mass parameter $\mu$).
  The right part shows the masses obtained from \texttt{Softsusy} via
  renormalization group running from the high-energy scale to the weak scale
  (the charged-Higgs-boson mass $m_{H^{\pm}}$, the gluino mass $m_{\tilde{g}}$,
   and the sbottom and stop masses $m_{\tilde{b}1}$, $m_{\tilde{b}2}$, $m_{\tilde{t}1}$, $m_{\tilde{t}2}$).}
\label{tableSPS}
\end{table}

In order to fix the MSSM parameters, we adopt the scenarios
Snowmass Points and Slopes (SPS) 1a and SPS4
\cite{Allanach:2002nj, Ghodbane:2002kg}, which are both derived from minimal
supergravity (mSUGRA).  While SPS1a is a typical mSUGRA scenario with a moderate
value of $\tan\beta$, $\tan\beta = 10$, the SPS4 point has a high value,
$\tan\beta=50$.
In the SPS4 scenario, we therefore expect large corrections from
$\tan\beta$-enhanced
contributions.  The input parameters for both scenarios are summarized in
Table~\ref{tableSPS}. They are defined at a high-energy unification scale.  We
use the program \texttt{Softsusy2.0.17} \cite{Allanach:2001kg} to perform the
renormalization group evolution from the high-energy scale to the weak scale and
to calculate the spectrum from the SPS input parameters. The resulting
charged-Higgs-boson and relevant SUSY-particle masses are also shown in
Table~\ref{tableSPS}.
The latter are defined in \texttt{Softsusy2.0.17} according to the modified
minimal-subtraction ($\overline{\rm DR}$) scheme of dimensional reduction.
They enter our core calculations only at one loop, so that a change of scheme
would induce a shift beyond NLO.
For the transition to the $\overline{\rm MS}$ scheme, such a shift would be
numerically insignificant \cite{Marquard:2007uj}.
In order to study the dependence on $\tan\beta$ and $\mhpm$, we
vary the high-scale input parameters and recalculate the full spectrum. While
$\tan\beta$ can be directly used as an input parameter to the spectrum
calculation, we adjust $m_0$ to obtain the desired value of $\mhpm$.
This also results in a change of the squark masses.

We are interested in a kinematical region where the center-of-mass energy
$\sqrt{s}$ is not sufficient to produce pairs of charged Higgs bosons,
i.e.\ $\sqrt{s}< 2m_{H^{\pm}}$, but sufficient to produce all
final-state particles on-shell, i.e.\  $\sqrt{s}> m_{H^{\pm}}+m_b+m_t$. For a
800~GeV collider, this implies that the interesting region lies in the range
$400 \GeV < \mhpm < 620 \GeV$.

\begin{figure}
\centerline{
\includegraphics[width=0.48\textwidth]{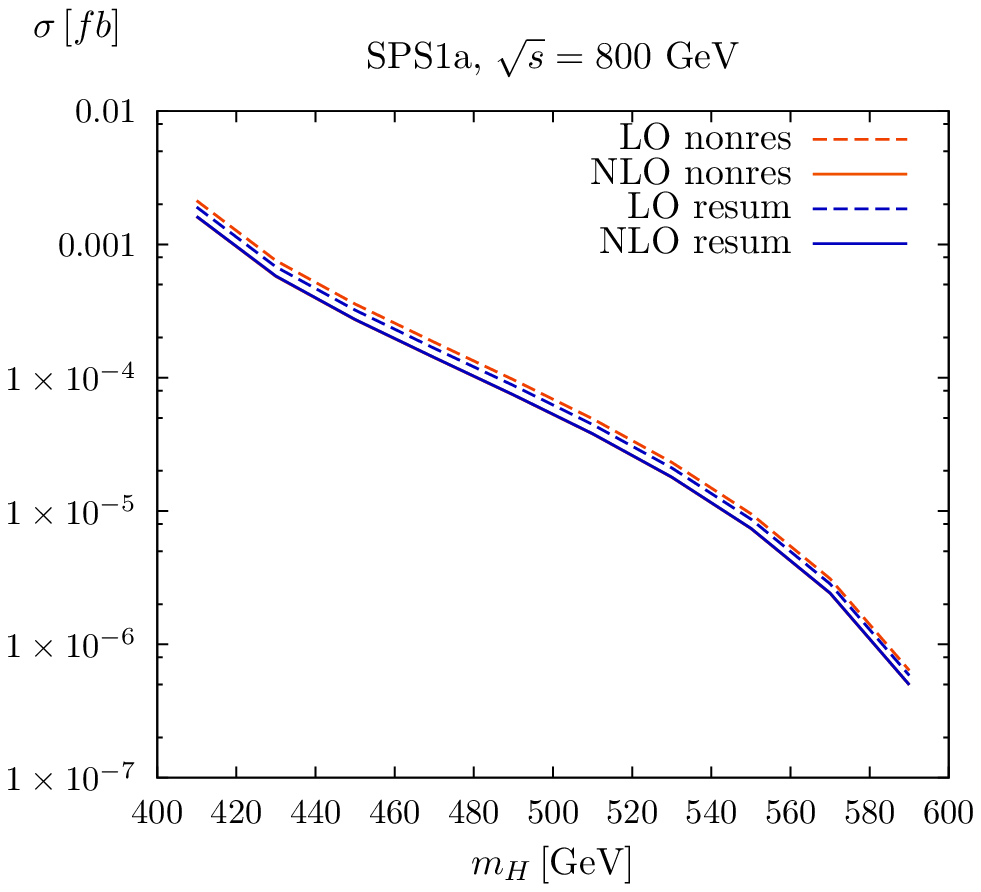}\qquad
\includegraphics[width=0.45\textwidth]{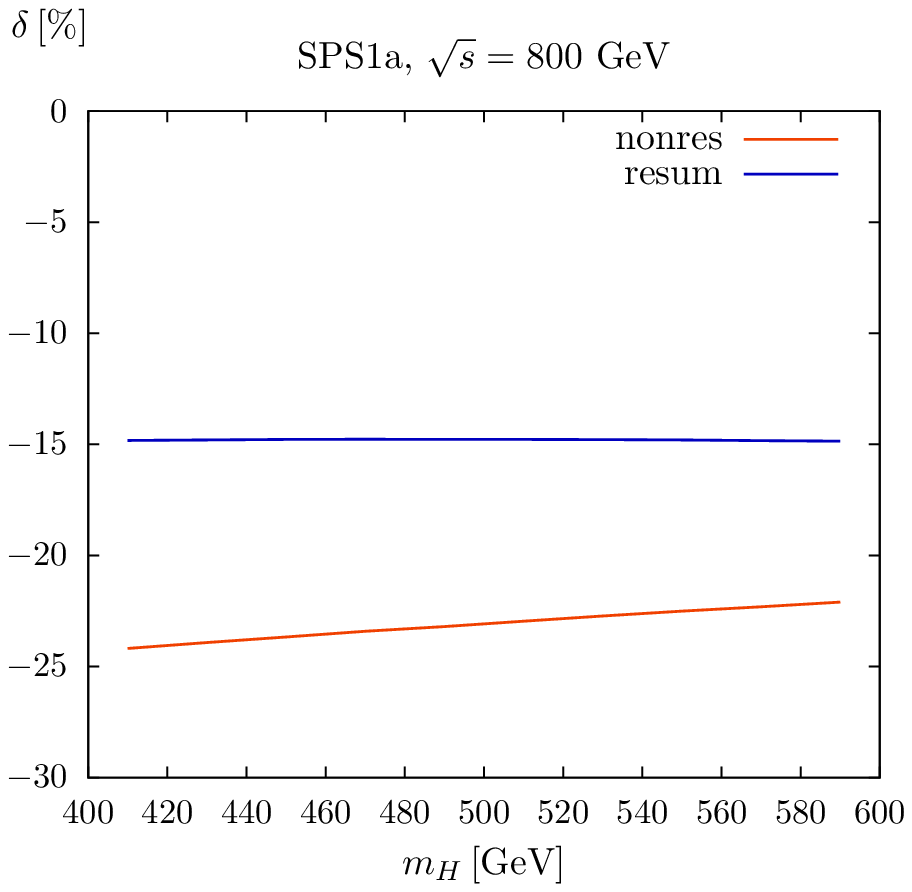}}
\caption{\em Total cross section $\sigma(e^+e^- \rightarrow t \bar{b} H^-)$ (left panel)
  and relative corrections $\delta$ in percent (right panel) depending on the
  charged-Higgs-boson mass $\mhpm$ for a center-of-mass energy of
  $\sqrt{s}=800$~GeV. The other MSSM parameters are taken to have their SPS1a values.
Both the nonresummed and resummed cross sections are considered.}
\label{fig:mhp-sps1a}
\end{figure}
\begin{figure}
\centerline{
\includegraphics[width=0.48\textwidth]{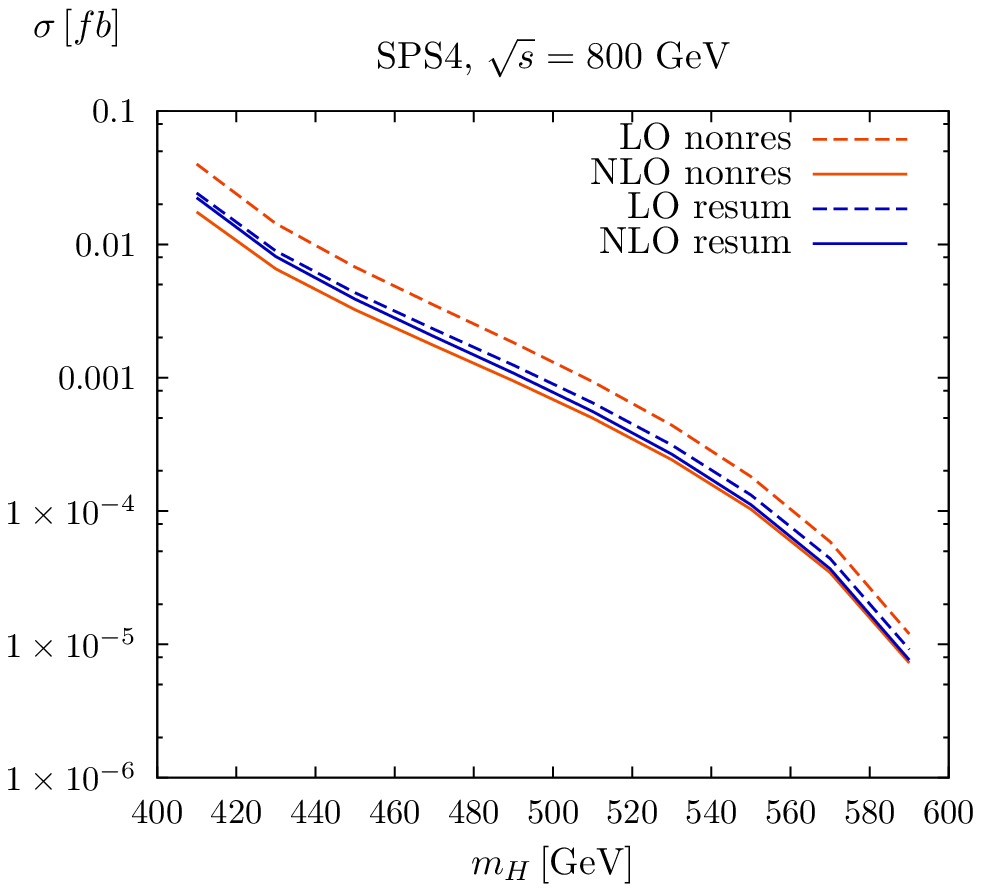}\qquad
\includegraphics[width=0.45\textwidth]{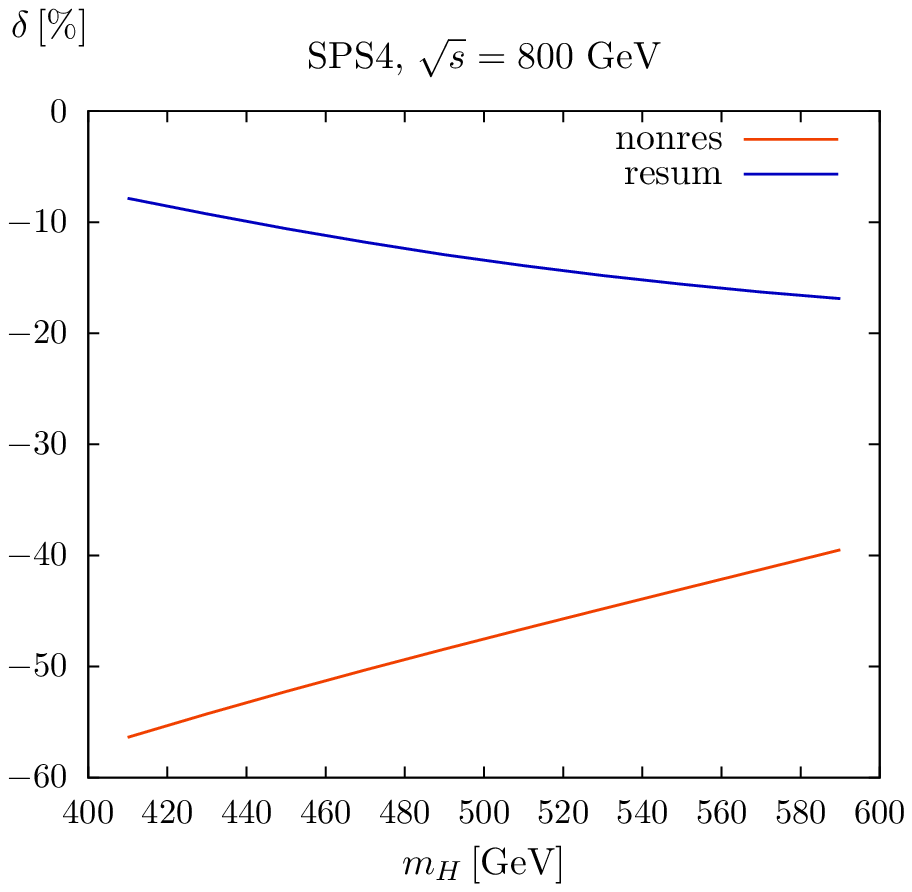}}
\caption{\em Total cross section $\sigma(e^+e^- \rightarrow t \bar{b} H^-)$ (left panel)
  and relative corrections~$\delta$ in percent (right panel) depending on the 
  charged Higgs-boson mass~$\mhpm$ for a center-of-mass energy of
  $\sqrt{s}=800$~GeV. The other MSSM parameters are taken to have their SPS4 values.
Both the nonresummed and resummed cross sections are considered.}
\label{fig:mhp-sps4}
\end{figure}
In Fig.~\ref{fig:mhp-sps1a}, the total cross section
$\sigma(e^+e^- \rightarrow t
\bar{b} H^-)$ is shown for the SPS1a scenario as a function of $\mhpm$
in the range $\sqrt{s}/2
< m_{H^{\pm}} < \sqrt{s}-m_b-m_t$. The cross
section falls rapidly with increasing distance from the pair-production
threshold at $\mhpm = \sqrt{s}/2 = 400 \GeV$.
Before resummation, the SUSY-QCD corrections range
from $-22\%$
to $-24 \%$ and exhibit only a weak dependence on $\mhpm$.
After resummation,
the residual corrections still amount to about $-15\%$, which
implies that the $\tan\beta$-enhanced contributions are not actually dominant
for these values of parameters.

Figure~\ref{fig:mhp-sps4} shows the $\mhpm$ dependence for the SPS4
scenario with $\tan\beta = 50$.
In this case, the SUSY-QCD corrections are much larger,
ranging from $-40\%$ to $-55\%$.
However, the bulk of these corrections is absorbed
by the resummation.  The size of the residual corrections
ranges from $-8 \%$ to $-17\%$
and is, therefore, similar to the SPS1a scenario.

\begin{figure}
\centerline{
\includegraphics[width=0.48\textwidth]{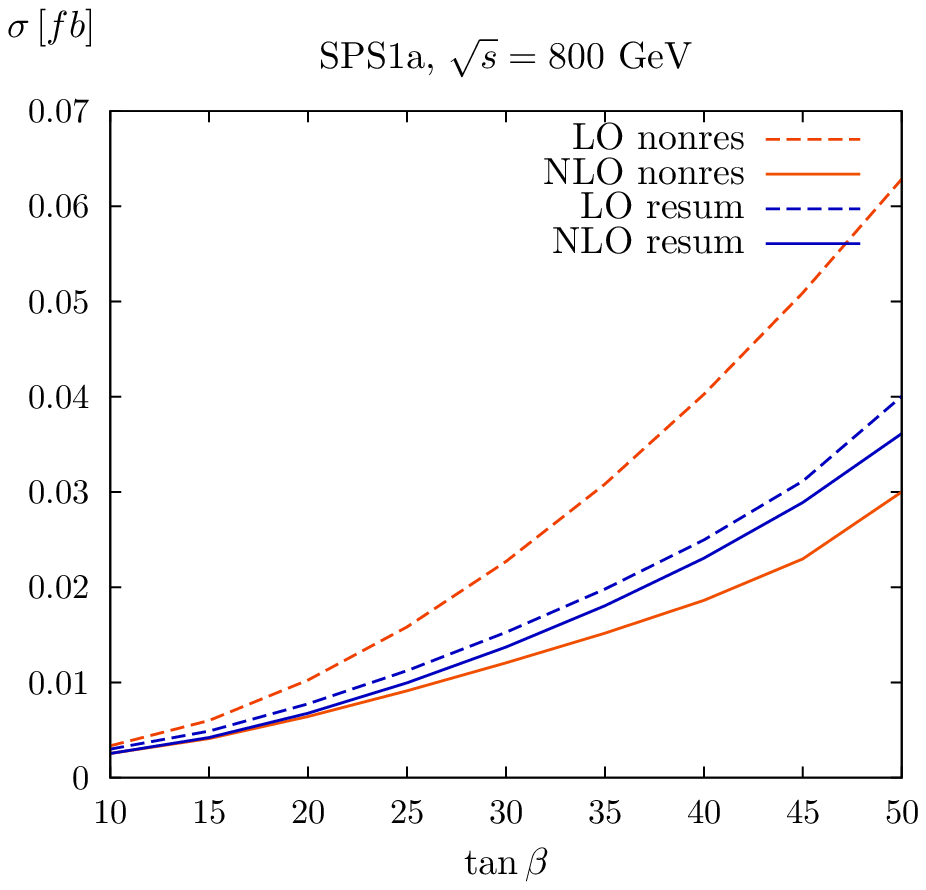}\qquad
\includegraphics[width=0.465\textwidth]{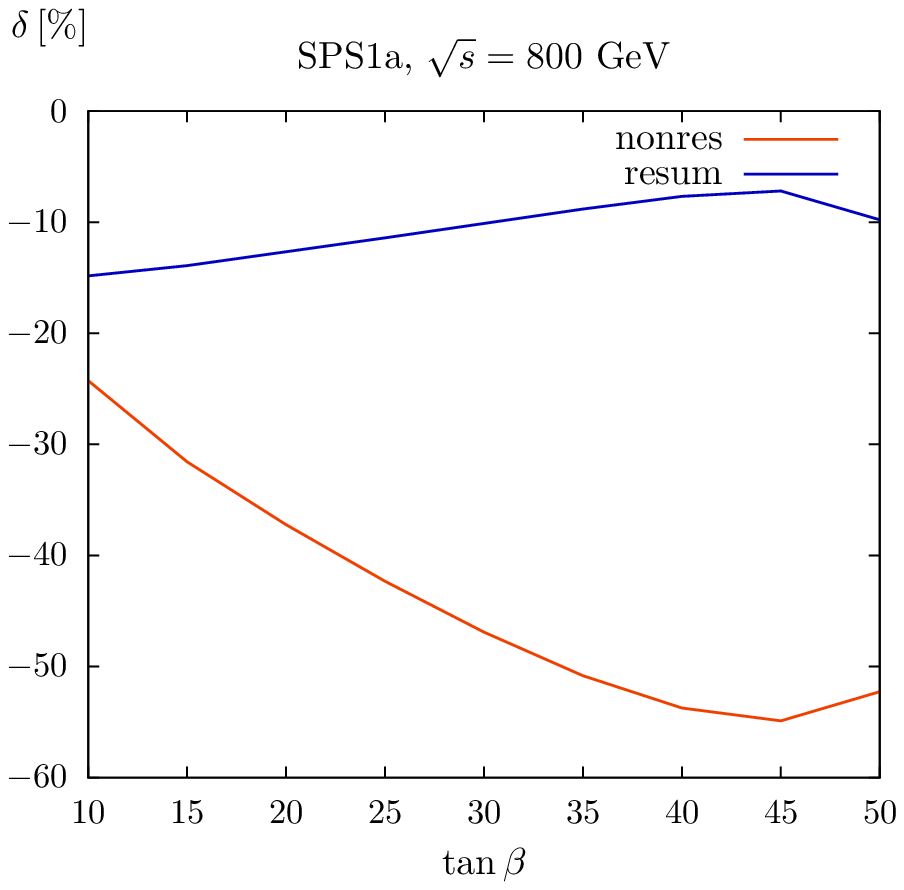}}
\caption{\em Total cross section $\sigma(e^+e^- \rightarrow t \bar{b} H^-)$ (left panel)
  and relative corrections~$\delta$ in percent (right panel) as a function of $\tan \beta$ for
  a center-of-mass energy of $\sqrt{s}=800$~GeV. The other MSSM parameters have
  their SPS1a values.
Both the nonresummed and resummed cross sections are considered.}
\label{fig:tanb}
\end{figure}
In Fig.~\ref{fig:tanb}, the $\tan\beta$ dependence of the total cross section
$\sigma(e^+e^- \rightarrow t \bar{b} H^-)$ is shown for the SPS1a scenario. The
cross section grows strongly with increasing value of $\tan\beta$. Without resummation,
the relative corrections also show a strong $\tan\beta$ dependence and range
from $-24\%$ at low value of $\tan\beta$ up to $-55\%$ at high value of
$\tan\beta$. Most of these
corrections are absorbed by resummation; the remaining corrections range from
$-10\%$
to $-15\%$ and show only a relatively weak dependence on $\tan\beta$.  Since
we change $\tan\beta$ by modifying the high-energy input parameters, we also
need to change $m_0$ accordingly to keep the low-energy value of $\mhpm$ at
its desired value.
This leads to somewhat higher squark masses at high values of $\tan\beta$, so that
the loop corrections are suppressed there.
Numerically, we find that the competing effects of the loop suppression due to
heavier squark masses and of the increase in the bottom Yukawa coupling lead to
a maximum in the relative corrections at $\tan\beta \approx 45$.
The difference between the nonresummed and resummed NLO cross sections
measures the size of the higher-order contributions included by performing the
resummation. For high values of $\tan\beta$, these contributions are of similar
size to the
residual SUSY-QCD corrections underlining the need for resummation in this
region.

We note that the analogous analysis of the SUSY-QCD corrections to the
associated production of a neutral Higgs boson with a $b\bar b$ pair in $e^+e^-$
annihilation with and without resummation of $\tan\beta$-enhanced contributions
yielded similar results \cite{Hafliger:2005aj}.

\subsection{Comparison of numerical vs~analytical loop integral evaluation}

\begin{table}[t]
\centerline{
  \begin{tabular}{cr@{.}l@{\qquad}r@{.}l}
    $m_{H^-}$ [\GeV]
    & \multicolumn{2}{c}{$\sigma_{\text{virt}}^{\text{BT}}\; [\fb]$}
    & \multicolumn{2}{c}{$\sigma_{\text{virt}}^{\text{ana}}\; [\fb]$} \\ \hline
    410 &  -5&154(7)$\cdot10^{-4}$  &  -5&148(4)$\cdot10^{-4}$ \\
    450 &  -8&46(1)$\cdot10^{-5}$   &  -8&462(6)$\cdot10^{-5}$ \\
    490 &  -2&245(3)$\cdot10^{-5}$  &  -2&245(2)$\cdot10^{-5}$ \\
    510 &  -1&129(2)$\cdot10^{-5}$  &  -1&1295(8)$\cdot10^{-5}$ \\
    550 &  -2&152(3)$\cdot10^{-6}$  &  -2&151(1)$\cdot10^{-6}$ \\
    590 &  -1&409(2)$\cdot10^{-7}$  &  -1&4094(9)$\cdot10^{-7}$ \\ \hline
  \end{tabular}}
  \caption{\em Contribution of the virtual corrections to the total cross
section
of the
    process $e^+ e^- \rightarrow t \bar{b} H^-$ for a
    center-of-mass energy of $\sqrt{s} = 800 \GeV$ using the BT and the
    analytical approaches.
The SUSY parameters are chosen according to the SPS1a scenario.}
  \label{tab:bt-cmp}
\end{table}

Let us now compare the BT method with the conventional analytical evaluation of
the loop integrals.
In~\tabre{tab:bt-cmp}, we compare the results for the virtual cross section
from Fig.~\ref{fig:mhp-sps1a} evaluated using the BT method and the analytical
loop integrals.
The remaining parameters are again chosen according to the SPS1a scenario.
Both methods use
the same number of points in the numerical integration performed by
\vegas, although the dimension of the integral is higher for the numerical
approach due to the additional Feynman-parameter integrations.
The runtimes of both approaches are also similar.  A comparison of the
results shows that the agreement between both methods is good, while the errors
are typically a factor 1.5 to 3 larger for the BT method. Detailed comparison
of the results from Figs.~\ref{fig:mhp-sps4} and \ref{fig:tanb} shows a
similar behavior.
\begin{figure}[t]
  \centerline{\includegraphics[width=0.49\textwidth]{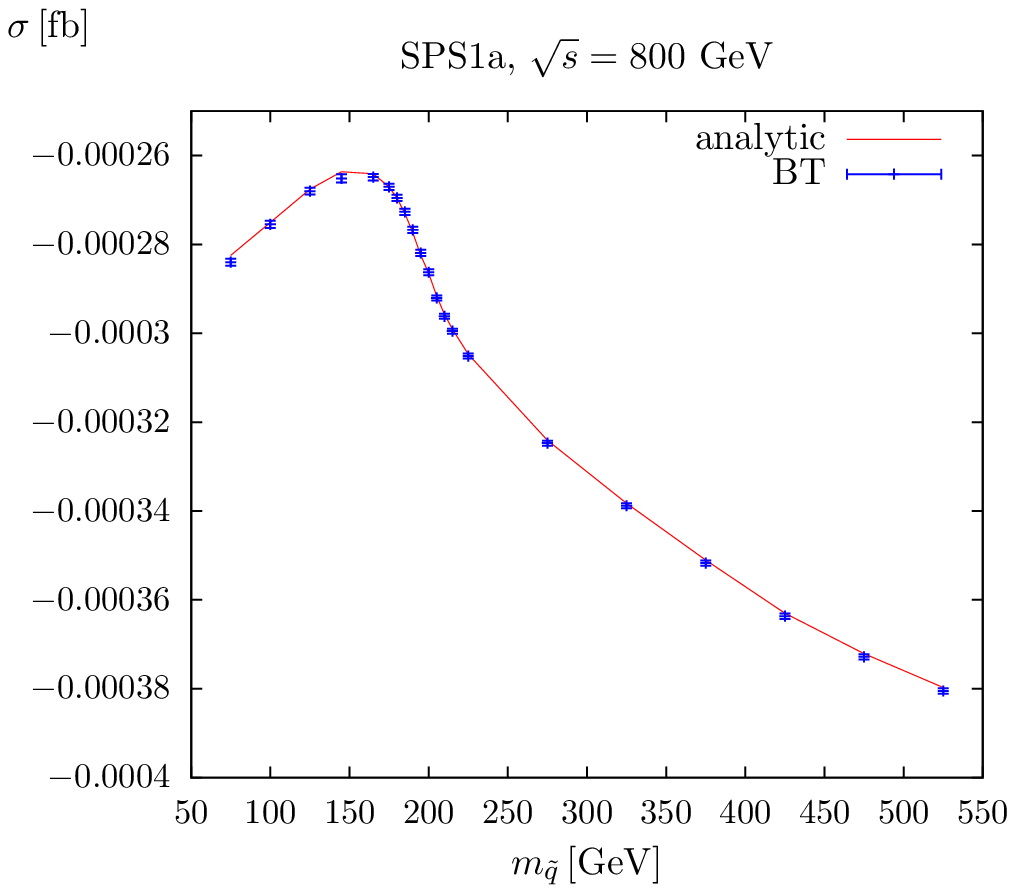}
\qquad
  \includegraphics[width=0.48\textwidth]{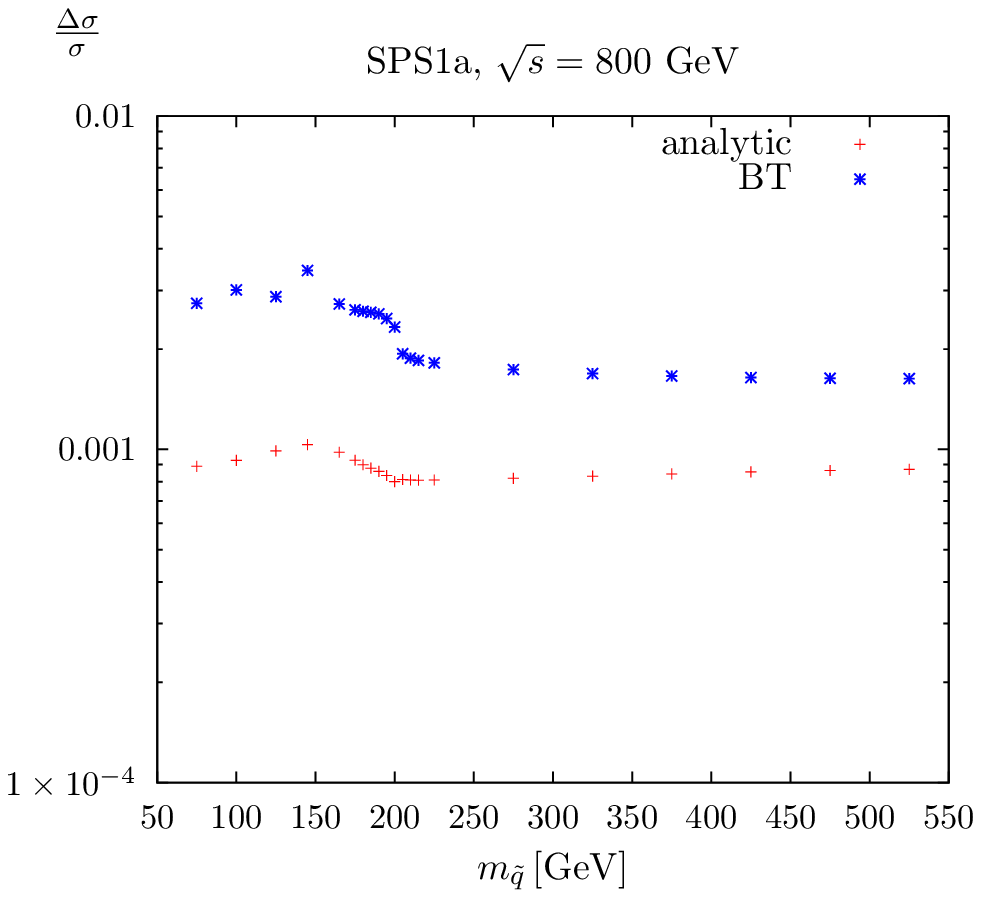}}
\caption{\em Contribution $\sigma_{\text{virt}}(e^+e^- \rightarrow t
  \bar{b} H^-)$ of the virtual corrections (left panel) to the
 total cross section and the integration error
  normalized to the virtual cross section (right panel) as functions of the common
  squark mass~$m_{\tilde{q}}$ for a center-of-mass energy of $\sqrt{s} = 800 \GeV$ using the BT
  and the analytical methods. The remaining SUSY parameters are chosen according
  to the SPS1a scenario.}
  \label{fig:bt-sigma-default}
\end{figure}

Evaluating single loop integrals using the BT method, we observe that the
errors and the runtime may depend sensitively on the presence
of thresholds of the particles in the loops. Below all thresholds, the
integrand is positive definite in Feynman-parameter space, so that the loop
integral is real and the numerical integration is very fast.
Once a threshold is crossed, the integrand becomes indefinite, the
integrals develop imaginary parts, and the deterministic integration
routine needs many more integrand evaluations to reach a reasonable
accuracy. In order to find out if this behavior carries over to the
combined phase-space and Feynman-parameter integration, we
calculate the virtual corrections keeping the center-of-mass energy fixed while
varying the squark masses. In this way, we can study the behavior in the
vicinity of
the various squark-squark thresholds in the loop diagrams.
Figure~\ref{fig:bt-sigma-default} shows the cross section and the
integration error for the virtual corrections as a function of a
common third-generation squark mass $m_{\tilde{q}} = m_{\tilde{t}_1} =
m_{\tilde{t}_2} = m_{\tilde{b}_1} = m_{\tilde{b}_2}$. For the
remaining parameters, we again use the SPS1a values.
If a common squark mass is used, the highest threshold in the loop diagrams
is the squark-squark threshold at $m_{\tilde{q}} =
\sqrt{s}/2 = 400 \GeV$ originating from diagrams
of the type shown in \figre{diags-virt}(a) and (e).
Cutting both squark lines in diagrams (b)
or (c), gives another squark-squark threshold at $m_{\tilde{q}} =
m_{H^\pm}/2 \simeq 200 \GeV$. The squark-squark threshold from
diagram (d) is at $m_{\tilde{q}} = s_{tb}/2$, which is
constrained by kinematics as $\sqrt{s_{tb}} < \sqrt{s} - m_{H^\pm} \simeq
400 \GeV$.
Here, $s_{tb}$ denotes the invariant mass square of the top-bottom system.
Since this is close to $m_{H^\pm}$ for our choice of parameters,
the virtual charged Higgs boson can become almost on-shell resulting in an
enhancement of diagram (d). The corresponding threshold below
$m_{\tilde{q}} \simeq 200 \GeV$ may appear to be dominating the
behavior of the total cross section in \figre{fig:bt-sigma-default}.

The error in the BT method shows a similar qualitative behavior to the error in
the analytical method, although it is a factor of 2 to 3 larger.
In particular, no
increase of the error below the threshold at $m_{\tilde{q}} = 400 \GeV$ can be
seen, and the increase in the error below $200 \GeV$ is only moderate.
The strong
variation of the errors across thresholds observed when calculating single
one-loop integrals in the BT method using a deterministic integration routine
is
almost completely absent when using the combined phase-space and
Feynman-parameter integration.



%
%
%
%
%
\section{Conclusion}

The discovery of charged Higgs bosons would be direct evidence of physics
beyond the SM. Since charged Higgs bosons are essential ingredients of all
SUSY models and, moreover, future colliders have the potential to
detect
them, it is worthwhile to study their phenomenology. If the
center-of-mass energy of the collider is not sufficient to produce
charged-Higgs-boson pairs, the process
$e^+ e^- \rightarrow t \bar{b} H^- +\text{c.c.}$ is of
particular interest.
We calculated the cross section of this process in the MSSM to NLO with regard
to the strong interaction, by complementing the known SM-QCD corrections
\cite{Kniehl:2002zz} by the SUSY-QCD ones.

As for the loop integrations, we deviated from the familiar analytical
approach, in which these loop-integrals are expressed in terms of logarithms
and dilogarithms.
Instead, we adopted the numerical approach of Ref.~\cite{Ferroglia:2002mz} based
on the BT theorem \cite{Tkachov:1996wh} and demonstrated its
feasibility for a multileg one-loop calculation such as ours.
We compared this alternative method
to the conventional analytical approach and found agreement within the
numerical
errors.
In contrast to previous applications of the BT method
\cite{Hollik:2005va, Passarino:2007fp,Actis:2008ug,Actis:2008ts},
we were tackling here for the first time a nontrivial phase-space integration
and demonstrated how it can be performed simultaneously with the BT
integrations.

Concerning the size of the corrections, we found significant SUSY-QCD
corrections ranging from~$-10\%$ to $-60\%$, which are of similar magnitude
to the purely gluonic QCD corrections \cite{Kniehl:2002zz}.
The bulk of SUSY-QCD corrections stem from
$\tan\beta$-enhanced contributions and can be absorbed by performing a
resummation of the bottom-quark Yukawa coupling. The residual corrections after
resummation are still sizeable, ranging from $-10\%$ to $-15\%$.


%
%
\section*{Acknowledgements}

The work of B.A.K. was supported in part by BMBF Grant No.\ 05~HT6GUA, by DFG
Grants No.\ KN~365/3--1 and No.\ KN~365/3--2, and by HGF Grant No.\ HA~101.

%
%


%
%


\end{document}